\documentclass[twocolumn,prl,aps,floatfix,floats]{revtex4}

\usepackage{graphicx}
\usepackage{dcolumn}
\usepackage{bm}
\usepackage{latexsym}
\usepackage{graphics}
\usepackage{dcolumn}
\usepackage{epsfig}
\usepackage{amssymb}
\usepackage{amsmath}
\usepackage{rotate}
\usepackage{color}
\newcommand{\gammap}{\dot{\gamma}}

\begin{document}

\title{Heterogeneities and dynamical phase transition in  flowing soft glassy materials .}
\title{Heterogeneities and dynamical phase transition in  flowing soft glassy materials .}
\title{A Boltzmann kinetic model for the cooperative rheology of soft glassy materials}
\title{A kinetic model for the cooperative rheology of soft glassy materials}
\title{Spatial cooperativity of soft glassy flows : a Boltzmann-like kinetic approach}
\title{From plastic events kinetics to spatial cooperativity in soft glassy rheology}
\title{As a dynamical phase transition}
\title{Kinetic Theory of plastic event dynamics and spatial cooperativity in soft glassy rheology}
\title{From Kinetic Theory of plastic event dynamics to spatial cooperativity in soft glassy rheology}
\title{Kinetic Theory of plastic events and spatial cooperativity in soft glassy rheology}
\title{A Kinetic Theory of plastic events and spatial cooperativity in soft glassy rheology}
\title{Soft glassy rheology derived from a coarse grained Kinetic Theory of plastic events}
\title{Kinetic Theory of plastic events, soft glassy rheology and spatial cooperativity}
\title{Kinetic Theory of plastic events and spatial cooperativity in soft glassy rheology}
\title{Soft glassy rheology from a coarse grained Kinetic Theory of plastic events dynamics}
\title{Cooperative rheology of jammed systems from a coarse grained Kinetic Theory of plastic events}

\title{Cooperative constitutive law for jammed flows from a coarse grained Kinetic Theory of plastic events}
\title{A kinetic theory of plastic flow in soft glassy materials}

\author{Lyd\'eric Bocquet$^{1}$\footnote{Corresponding author: {\tt lyderic.bocquet@univ-lyon1.fr}}, Annie Colin$^{2}$, Armand Ajdari$^{3}$}
\affiliation{$^{1}$Laboratoire PMCN, Universit\'e Lyon 1, Universit\'e de Lyon, UMR CNRS 5586, 69622 Villeurbanne, France\\ $^{2}$LOF,
Universit\'e Bordeaux 1, UMR CNRS-Rhodia-Bordeaux 1 5258, 33608
Pessac cedex, France\\ $^{3}$Laboratoire de Physico-Chimie Th\'eorique, UMR CNRS-ESPCI 7083, 10 rue Vauquelin, 75231 Paris Cedex 05, FRANCE}



\date{\today}
\begin{abstract}
A kinetic model for the elasto-plastic dynamics of a flowing jammed material is proposed, which takes the form of a non-local -- Boltzmann-like --  kinetic equation for the stress distribution function. Coarse-graining this equation yields a non-local constitutive law for the flow, introducing as a key dynamic quantity the local rate of plastic events. This quantity, interpreted as a local fluidity, is spatially correlated, with a correlation length diverging in the quasi-static limit, {\it i.e.}Ê  close to yielding. We predict finite size effects in the flow behavior, as well as the absence of an intrinsic local flow curves. These features are supported by recent experimental and numerical observations.
 \end{abstract}

\maketitle
%
%


Soft amorphous materials such as foams, emulsions, granular systems or colloidal suspensions 
display complex flow properties at high enough concentrations, intermediate between that of a solid and a liquid:
 at rest they behave like an elastic solid, but are able to flow ``like a liquid'' under sufficient applied stress \cite{coussot,dennin,vanhecke,Besseling}.
This mixed fluid/solid behavior occurs above a threshold volume fraction 
associated with the appearance of a yield stress $\sigma_d$. 
The yielding behavior makes such systems particularly interesting for applications -- from tooth paste, coatings to cosmetic and food emulsions --, but fundamentally difficult to describe \cite{Sollich:97,Hebraud:98,Fuchs}.
Furthermore, it has been recognized over the recent years that this yielding behavior is, in most cases, associated with peculiar {\it spatial} features. This takes the form of inhomogeneous flow patterns, such as shear-bands
\cite{coussot,dennin,vanhecke,varnik}, or cooperativity in the flow or deformation response \cite{Hebraud:97,Dauchot:05,Goyon:08,Hatano,Leonforte,Picard,Pouliquen2,deboeuf}, potentially associated with non-locality in the constitutive rheological law \cite{Goyon:08} and dependence of the flow
on the nature of the boundaries \cite{Besseling,Seb}. While such features appear to be generic 
for this class of materials, suggesting a underlying common flow scenario, a consistent
framework linking the global rheology to the local microscopic dynamics is still lacking.


In this paper, we present a kinetic elasto-plastic (KEP) model, which aims at 
constructing such a link between the microscopic and the macroscopic scales. 
Starting from a kinetic 
elasto-plastic  description of the dynamics, 
we derive systematically a (non-local) generic constitutive law for the flow, obtained by
coarse-graining the microscopic 
spatio-temporal dynamics. The predictions of the KEP model
will be shown to 
capture most features of the rheology of yield fluids, and in particular the recent experimental demonstration of cooperativity in the flow behavior of jammed emulsions \cite{Goyon:08}.

The KEP model, which is detailed below, is based on the generic picture which has emerged recently for the dynamics of soft glassy materials \cite{Lemaitre,Picard,Schall}.
In these materials, flow occurs through a succession of elastic deformations and {\it localized} plastic rearangements associated with a microscopic yield stress. 
These localized events induce long range elastic modifications of the stress over the system, thereby creating long-lived fragile zones where flow occurs. 
Flow in these systems is thus highly cooperative and spatially heterogeneous: a dynamically active region will induce agitation of its neighbours and thus a locally higher rate of plastic rearrangements. 
Correlations between plastic events are accordingly expected to exhibit a
complex spatio-temporal pattern \cite{Picard}.

{\it The KEP model --} Describing these complex dynamical processes is a formidable task, and to get further insights, we propose on purpose a schematic model, relying on a few simplifying assumptions.
To this end, the KEP model extends
on an approach first proposed by H\'ebraud and Lequeux (HL) \cite{Hebraud:98}, by describing spatial interactions between plastic events:
the sample is divided into elementary blocks $i$ of size $a$ (typically the size of individual particles), 
carrying a scalar shear-stress $\sigma_i$ \footnote{This scalar assumption could be removed along the lines proposed by Cates and Sollich \cite{Cates:04} for the SGR model.}, and the system is described in terms of the block stress distribution $P_i(\sigma, t)$.
The latter evolves via three mechanisms: an {\it elastic response}, under an externally imposed shear rate $\gammap_i^o$; a
stress relaxation due to {\it local plastic events};
 the modification of stress due to the plastic events {\it occuring in other blocks}, transmitted spatially via elastic interactions. 
 Various simplifying assumptions are made to describe these processes. First, the local plastic events will be assumed to occur above a a local threshold value of the stress $\sigma_c$. 
%
The elastic propagation of the shear stress is captured using the stress-stress elastic propagator $\Pi_{i,j}$ \cite{Picard}, relating the stress relaxed at a block $i$ due to the occurence of a localized plastic event in {\it another block} $j$:
$\delta \sigma_i=\Pi_{i,j}\cdot\delta \sigma_j$  where  $\delta \sigma_j$ is
the relaxed stress at block $j$ (we assume here a full stress relaxation $\delta \sigma_j=-\sigma_j $). Furthermore, in order to get a closed kinetic equation, we propose a decoupling of the plastic-event dynamics, in the same spirit as the Boltzmann Stosszahlansatz.  Finally, as a first approach, convection is neglected.
 Altogether, within these simplifying hypothesis, the KEP equation for $P_i(\sigma,t)$ takes a Boltzmann-like form
\begin{alignat}{2}
\partial_t P_i(\sigma,t)=-G_{o}\gammap_i \partial_\sigma P_i (\sigma,t) -\frac{\Theta(\vert \sigma \vert -\sigma_c)}{\tau}P_i (\sigma,t)\notag\\
+\Gamma_i (t)\delta(\sigma)  +\mathcal{L} (P,P)  
\label{HL}
\end{alignat}
with $G_{o}$ is the elastic modulus, $\Theta $ the Heaviside function.
The non-local Boltzmann like operator $\mathcal{L} (P,P) $ is defined according to :
\begin{alignat}{2}
\mathcal{L}(P,P)=\sum_{j\neq i} \int  d\sigma'\, \frac{\Theta(\mid \sigma' \mid -\sigma_c)}{\tau} \notag\\
\left[P_j(\sigma',t)P_i(\sigma+\delta \sigma_i,t)-P_j(\sigma')P_i(\sigma)\right]
\end{alignat}
with $\delta \sigma_i = \Pi_{i,j} \delta \sigma_j=- \Pi_{i,j} \sigma'$ and $1/\tau$ acounts for the relaxation rate when the stress
is larger than $\sigma_c$.
$\mathcal {L}(P, P)$ 
describes
the gain and loss contributions for the probability $ P_i(\sigma, t)$
due to events occuring in other blocks, in full analogy with the Boltzmann equation.  
The rate of plastic events, $\Gamma_i (t)$, entering Eq. (\ref{HL}), is defined as 
%
\begin{alignat}{2}
\Gamma_i (t)= \int \frac{\Theta(\mid \sigma' \mid -\sigma_c)}{\tau}P(\sigma')d\sigma',
\label{fluidity}
\end{alignat}

In its above form, the KEP equation remains difficult to solve analytically.
To proceed further, we formally expand the Bolztmann operator $\mathcal {L}(P, P)$ for small stress variations $\delta \sigma$ and retain 
only the first terms of the expansion. The further simplification $\delta \sigma_j \approx- \sigma_c$  (valid for small $\gammap$) is also made.
This simplifies Eq. (\ref{HL}) to a Fokker-Planck equation:
\begin{alignat}{2}
\partial_t P_i(\sigma,t)=-G_{o}\gammap_i\partial_\sigma P_i(\sigma,t)-\frac{\Theta(\mid \sigma \mid-\sigma_c)}{\tau}P_i(\sigma,t) \notag\\
+\Gamma_i(t)\delta(\sigma) +D_i\partial^2_{\sigma^2} P_i(\sigma,t)\label{bloc1}
\end{alignat}
In this equation,  
$\gammap_i$ 
is the local shear-rate ($\gammap_i=\gammap_i^o+ \frac{1}{2}  \sum_{j \neq  i} \Pi_{ij} \sigma_c \Gamma_j$)
and
the coefficient $D_i$ quantifies what appears as a {\it stress diffusion} induced by the occurence of plastic events. 
A key result is that stress diffusion is related to the rate of plastic events over the whole system via the self-consistency relationship: 
$D_i=\frac{1}{2} \sum_{j \neq  i} \Pi_{ij}^2\sigma_c^2\, \Gamma_j$,
therefore making the Fokker-Planck equation non-linear.

Coming back to continuous spatial variables, a closed system of equations is obtained for the local stress diffusion
$D({\mathbf r},t)$, rate of plastic events $\Gamma({\mathbf r},t)$ and stress distribution $P(\sigma,{\mathbf r},t)$. Eq.~(\ref{bloc1}) keeps the same form (with $i\rightarrow {\mathbf r}$), while a small slope approximation of the self-constistency equation for $D$ 
provides 
a non-local relationship between stress diffusion and rate of plastic events:
\begin{alignat}{2}
D({\mathbf r},t)=m\, \Delta \Gamma({\mathbf r},t) +\alpha\, \Gamma({\mathbf r},t)
\label{bloc5}
\end{alignat}
with $\Delta$ the spatial Laplacian.
In this equation, two key parameters have been introduced: a {\it coupling parameter} $\alpha$, here defined as $\alpha=\sigma_c^2 \sum_{i\ne j} \Pi_{i,j}^2$; and an inhomogeneity parameter $m= a^2 \sigma_c^2 \Pi_{nn}^2$, with $\Pi_{nn}$ the nearest neighbour (block-to-block) propagator.
In the following we will make use of dimensionless variables, 
$\tilde{t}={t}/{\tau}$, $\tilde{r}={r}/{a}$, $\tilde{\sigma}={\sigma}/{\sigma_c}$, $ \tilde{\gammap}_{\rm loc}=\gammap_{\rm loc}\, {G_o \tau }/{\sigma_c}$,
$\tilde{m}= m/a^2\sigma_c^2$, $ \tilde{\alpha}=\alpha/\sigma_c^2 $,  $\tilde{\Gamma}=\Gamma \tau$,   and $\tilde{D}={D \tau}/{\sigma_c^2}$, but will drop the $\tilde{ }$ to simplify notations.

If inhomogeneities are neglected [{\it e.g.}, putting $m=0$ in Eq.~(\ref{bloc5})], the above set of equations reduce exactly to the HL description in Ref. \cite{Hebraud:98}. Anticipating on the discussion below, a key result which emerges from the HL description is that it predicts a {\it jamming
transition} below a threshold (dimensionless) coupling parameter $\alpha<\alpha_c={1\over 2}$, associated with the building-up of a macroscopic {\it dynamic yield
stress},  $\sigma(\gammap\rightarrow 0)= \sigma_d$. As shown in Ref. \cite{Hebraud:98}, $\sigma_d \propto (\alpha_c -\alpha)^\beta$,Ê with $\beta=1/2$ and this dynamic yield stress thus 
quantifies the distance to the jamming transition \cite{note}.
In the following we shall focus on the jammed state, as defined by a non-vanishing $\sigma_d$. 
\vskip0.2cm


%

{\it Constitutive flow rules --} 
In the stationnary state, the Fokker-Planck equation, Eq. (\ref{bloc1}), can be solved analytically \cite{Hebraud:98} to give an explicit expression for $P(\sigma,{\mathbf r})$. One then deduces
the local averaged stress $\bar\sigma ({\mathbf r})=\int d\sigma'\, \sigma'P(\sigma',{\mathbf r})$ and the local rate of plastic events $\Gamma({\mathbf r})$ (from the normalization condition for $P$). This
provides explicit expressions for these quantities in terms of the diffusion coefficient $D({\mathbf r})$ and local shear rate $\gammap({\mathbf r})$.
While their general expression is
rather cumbersome, they simplify considerably in the quasistatic limit ($\gammap\rightarrow 0$) and close to the jamming point,
{\it i.e.} small $\sigma_d$.
Choosing the plastic rate $\Gamma$ as the key variable, one obtains the following expressions in this regime: 
$\bar\sigma ={(6 \Gamma)}^{-1} \times \gammap$ and
$D-\alpha \Gamma = a_1 \sigma_d (\sigma_d-\bar\sigma)\Gamma+a_2 \Gamma^{3/2} + {\cal O}(\Gamma^{2})$,
with $\sigma_d$  the dynamic yield stress introduced above, and $a_1$, $a_2$ two numerical constants \cite{note}.
In the following we define $f=6 \Gamma$ as the {\it fluidity}: the latter naturally emerges as intimately linked to the rate of plastic events.

Together with the self-consistency relationship Eq. (\ref{bloc5}), relating $D$ to $\Gamma$,
these expressions provide a closed set of equations. A further linearization
allows to rewrite this set in the physically meaningful form:
\begin{eqnarray}
\bar\sigma =\frac{1}{f} \times \gammap\notag\\
 \bigtriangleup f -\frac{1}{\xi^2}(f-f_b)=0
\label{bloc9}
\end{eqnarray}
In this equation we have introduced a ``{\it bulk fluidity}'' $f_b(\bar\sigma)$: $f_b(\bar\sigma)=6 (\frac{a_1 \sigma_d}{a_2})^2 (\bar\sigma-\sigma_d)^2$ for $\bar\sigma> \sigma_d$ and 0 otherwise;
and a fluidity {\it correlation length} $\xi(\bar\sigma)$:
 $\xi=\sqrt\frac{2m}{a_1(\bar\sigma-\sigma_d)}$ for $\bar\sigma  > \sigma_d$ and $\xi=\sqrt\frac{m}{a_1(\sigma_d-\bar\sigma)}$ for $\bar\sigma<\sigma_d$. 
These coupled equations constitute the {\it non-local constitutive flow rules} which emerge from the KEP model, and are the central result of this work.


The bulk fluidity $f_b(\bar\sigma)$  is the value of the fluidity obtained in absence of non local terms, as obtained in the HL model: as can be easily verified, it predicts a Herschel-Bulkley expression for the flow rule for low shear rates, with $\sigma_d$ as the dynamic yield stress: $\bar\sigma(\gammap)=\sigma_d + A \gammap^{n}$, with $n=1/2$ and $A$ a constant depending on $\alpha$. A further key result from
Eq. (\ref{bloc9}) is the {\it non-local nature} of the 
flow curve, which  introduces a ``{\it flow cooperativity length}'' $\xi$. Physically, $\xi$ quantifies the spatial spreading of the plastic activity  due to the non-local elastic relaxation over the system. Interestingly, the correlation length diverges at the dynamical yield stress according to $\xi\propto \vert\sigma-\sigma_d\vert^{-1/2} \propto \gammap^{-1/4}$, in agreement with recent
numerical simulations \cite{Hatano}.

We emphasize that the non-local flow rule predicted by the KEP model, Eqs.~(\ref{bloc9}), is 
formally identical 
to the cooperative rheology introduced recently to account for the flow of confined jammed emulsions \cite{Goyon:08}.



 \begin{figure}[t]
\begin{center}
\includegraphics [width=8 cm,height=6cm] {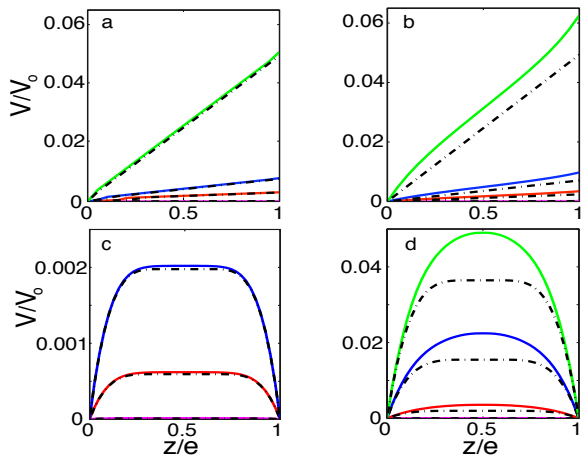}
\end{center}
\caption{{\it (Top)}  Dimensionless velocity profiles of a jammed material under pure shear flow in a slit of width $e=40\,a$ 
($\alpha=0.45<\alpha_c$). The full lines are deduced from Eq.~(\ref{bloc10}) and the dashed lines from the bulk model ($m=0$ {\it i.e.} $\xi_0=0$). {\it (a)}: $\xi_0\approx a$; from bottom to top  $\bar\sigma=1.5, 3, 4, 7$ $ \sigma_d$. 
{\it (b)}: same with $\xi_0 \approx 8 a$. 
{\it (Bottom)} Pressure driven flows 
{\it (c)}: $\xi_0\approx a$; 
From bottom to top, the lines correspond to various shear stress at the wall $\sigma_w=1.5, 3, 4$ $\sigma_d$. 
{\it (d)}: same with $\xi_0 \approx 8 a$. $\sigma_w=1.5, 4,7,9\sigma_d$. 
The characteristic flow velocity is defined as $V_0={\sigma_c e}/{G_o \tau}$.
As a boundary condition, we chose the wall fluidity as  $f_w=2\, f_b(\bar\sigma)$.}
\label{a}
\end{figure}

{\it Couette and Pressure-driven flows --}
Let us now discuss the solution of this rheological model in various geometries. One expects non-local effects to emerge in the flow behavior and
we introduce a characteristic length $\xi_0$, 
which we define as $\xi_0=\xi(\sigma_d + \delta \sigma)$, with the somewhat arbitrary choice $\delta\sigma= {1\over 2}\sigma_d$. Non-local effects are
expected when the size of the system, say $e$, is comparable to the correlation length $\xi_0$. 
The flow behavior also requires boundary conditions for the fluidity at the confining walls, which -- in line with experimental results \cite{Goyon:08} -- we will assume here to be a given function of the stress at the walls, $f_w\equiv f_w(\sigma_w)$. 

For a planar Couette cell made of two parallel walls separated by a distance $e$,  the mean shear stress $\bar\sigma({\mathbf r})$ is spatially homogeneous $\bar\sigma=\sigma_o$ and the resolution of Eq. (\ref{bloc9}) is straightforward in this geometry. 
This provides the expression of $\gammap$:
 \begin{alignat}{2}
\gammap(z)=\bar\sigma\cdot\left(f_b+(f_w-f_b)\frac{\cosh[(z-{e/2})/ \xi(\sigma_o)] }{\cosh[e/2\xi(\sigma_o)] }\right)
\label{bloc10}
\end{alignat}
 where $z$ is the distance from the bottom plate. Velocity profiles are deduced by integration (assuming here no-slip boundary condition at the walls).
 Figs. \ref{a}(a)-(b) show the resulting velocity profiles for various characteristic lengths $\xi_0$. 
In the pressure driven (Poiseuille) geometry, the stress varies spatially in the confined channel according to $\bar\sigma=\nabla P (z-{e\over 2})$,
with $\nabla P$ the constant pressure gradient along the slit. The constitutive laws, Eqs.~(\ref{bloc9}), are integrated numerically for various
$\nabla P$ and the
resulting velocity profiles are displayed in Figs. \ref{a}(c)-(d).

In both Couette and Poiseuille geometries, the flow profiles deduced from the non-local constitutive rules depart strongly from the ``bulk'' prediction ({\it i.e.} without non-local effects), as soon as the characteristic length $\xi_0$ compares with the confinement 
 [note however that in Figs. \ref{a}(b)-(d) $\xi_0 $ is only $\xi_0\approx 0.2\, e$].  Furthermore the effect is more pronounced for the Poiseuille geometry, 
due to the spatial inhomogeneity of the stress map which indeed amplifies the non-locality effect.
%

\begin{figure}[tb]
\begin{center}
\includegraphics [width=8 cm] {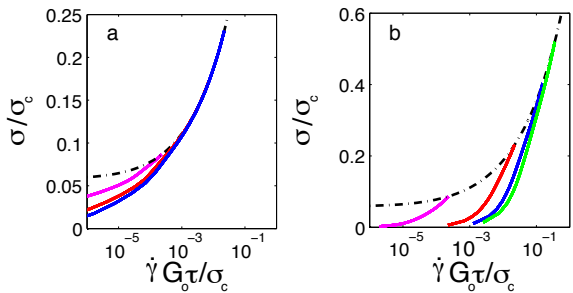}
\end{center}
\caption{Local flow curves $\sigma(z)$ versus $\gammap(z)$ extracted from the velocity profiles in the pressure driven flows of Figs. \ref{a}(c)-(d),
{\it i.e.} characterized by two different non-locality parameter {\it (a)}: $\xi_0\approx a$; {\it (b)}: $\xi_0\approx 8 a$.
The dashed line is the bulk flow curve (without non-locality).
As expected, deviations are more pronounced for the stronger non-locality.
Note the difference in vertical scales.}
\label{b}
\end{figure}

An alternative way of exhibiting non-locality is to plot the {\it local flow curve}: $\bar\sigma(z)$ versus $\gammap(z)$. Fig.~\ref{b} shows the result of such a plot for the Poiseuille geometry. As evidenced on these curves, the existence of non-locality (finite $\xi_0$) results in a {\it multivalued} local flow curve, departing from the bulk prediction:
different values for the shear rate $\gammap$ are obtained for the same value of the stress, obtained here for different pressure gradients $\nabla P$.
A similar multivalued behavior is also obtained upon varying the confinement $e$. 
In other words cooperativity  induces finite-size effects
in the flow of the jammed material which occurs for confinements $e \sim \xi_0$.
Furthermore as seen on the local flow curve, Fig.~\ref{b}, non-local effects do suppress 
the yielding behavior of the fluid. 

{\it Discussion --}Ê
Altogether these cooperative flow behavior predicted by the KEP model are in very good agreement with the recent experimental results for the flow of jammed emulsions in microchannels Ê\cite{Goyon:08}. In these experiments, the flow profiles were found to depart from the bulk prediction for confinements $e$ typically smaller than a few tens of droplet diameters. Furthermore,  a constitutive law similar to
Eq. (\ref{bloc9}) was able to rationalize all experimental results, with  a cooperativity length scales of the order of several droplets diameters. One difference however is that
no dependence of the cooperativity length on shear-rate was reported experimentally, even though the flow behavior in the quasi-static limit is difficult to access experimentally and would certainly require a further specific investigation. Furthermore similar cooperativity effects  are reported in granular
flows \cite{deboeuf,Pouliquen2,Hatano}, as well as in numerical simulations of deformation of amorphous materials \cite{Leonforte}.

Another prediction of the non-local constitutive law is the strong impact of boundary effects on the flow. As can be seen {\it e.g.}Ê
in Eq. (\ref{bloc10}), the flow profile within the cell is influenced by the wall fluidity. The latter is expected to depend on surface properties,
{\it e.g.} roughness : a smooth wall is indeed expected to induce a smaller wall fluidity as compared to a rough wall, which in turns will modify the shape of the flow profile in the material. The influence of boundary roughness  on the
flow is indeed observed experimentally in various systems \cite{Goyon:08,Besseling,Seb} and would definitely desserve a more systematic investigation.

At a more formal level, it is interesting to note that 
 the solution of Eq.~(\ref{bloc5}) is the minimum of the square gradient 'free-energy':
$\Omega({\Gamma})= \int d{\mathbf r}\, \frac{m}{2} (\nabla  \Gamma ) ^2+\omega (\Gamma,\bar\sigma)$, 
with 
$\omega (\Gamma,\bar\sigma)=\frac{1}{2} a_1\sigma_d (\sigma_d-\bar\sigma)\Gamma^2+\frac{2}{5}a_2\Gamma^{5/2}+O(\Gamma^4)$ in the limit of small $\Gamma$. 
This equation is analogous to a Landau expansion close to a second order phase transition, with the dynamic yield stress $\sigma_d$ as critical point. The rate of plastic events $\Gamma$, {\it i.e.} the fluidity, plays the role of the (dynamic) order parameter. 

Beyond the formal analogy,  this suggests a interesting alternative point of view for flow inhomogeneities.
While the present scenario predicts flow inhomogeneities characterized by a cooperativity length scale, in line with results for dense emulsions, a ``true'' shear-banding would merely correspond to a first order phase transition scenario:  {\it i.e.} the spatial coexistence between two states of different fluidity for the same shear stress. 
Recent experimental findings have
connected shear-banding to the existence of attractive interactions between particles, thereby inducing a flow-structure coupling in the material
\cite{Becu:06}. 
The KEP description does not account for these features and
it would be therefore interesting to include local structure variables in the description in order to capture such couplings.



{\it Conclusions --} 
In conclusion, we have derived a non-local constitutive equation for the flow of jammed systems from a `microscopic' kinetic elasto-plastic model.
The resulting description suggests the cooperative nature of the flow, in full agreement with recent experimental findings \cite{Goyon:08}.
Furthermore, this framework puts forward the role of the fluidity as a dynamical order parameter characterizing the flow, and here defined as the local rate of plastic events in the material. Since one expects plastic events to trigger local velocity fluctuations, $\langle\delta v^2\rangle$, the latter quantity could provide an indirect measure of the fluidity, in line with granular hydrodynamics approaches \cite{granu}.
It is interesting to note that similar observations of non-locality have been reported in granular flows close to the jamming transition \cite{Pouliquen2,deboeuf}, suggesting further universal characteristics.

{\it Acknowledgements --} LB thanks J.-L. Barrat, B. Andreotti, D. Durian and
P. Sollich for interesting discussions. LB and AC acknowledge support from ANR, program SYSCOMM.

\end{document}